\documentclass[twocolumn,showkeys,aps,prb,showpacs]{revtex4-1}
\UseRawInputEncoding
\usepackage{graphicx}
\usepackage[CJKbookmarks,dvipdfm,colorlinks,linkcolor=red,citecolor=blue]{hyperref}

\begin{document}

\title{Switching  Rashba spin-splitting  by reversing  electric-field direction  }

\author{San-Dong Guo$^{1}$, Jing-Xin Zhu$^{1}$, Hao-Tian Guo$^{1}$, Bing Wang$^{2}$, Guang-Zhao Wang$^{3}$ and Yee-Sin Ang$^{4}$}
\affiliation{$^1$School of Electronic Engineering, Xi'an University of Posts and Telecommunications, Xi'an 710121, China}
\affiliation{$^2$Institute for Computational Materials Science, School of Physics and Electronics,Henan University, 475004, Kaifeng, China}
\affiliation{$^3$Key Laboratory of Extraordinary Bond Engineering and Advanced Materials Technology of Chongqing, School of Electronic Information Engineering, Yangtze Normal University, Chongqing 408100, China}
\affiliation{$^4$Science, Mathematics and Technology (SMT), Singapore University of Technology and Design (SUTD), 8 Somapah Road, Singapore 487372, Singapore}
\begin{abstract}
The manipulation of the Rashba spin-splitting is crucial for the development of nanospintronic technology.
Here, it is proposed that the Rashba spin-splitting can be turned on and off by reversing  electric-field direction.
By the first-principle calculations,  our proposal  is illustrated by a concrete example of Janus monolayer RbKNaBi.
The designed RbKNaBi possesses dynamical, thermal and mechanical stability, and is a large-gap quantum spin Hall insulator (QSHI) with Rashba spin-splitting  near
the Fermi level.  A small built-in electric field is predicted due to very small  electronegativity difference between the bottom and top atoms, which is very key to switch Rashba spin-splitting  through the experimentally available electric field intensity.
Due to out-of-plane  structural asymmetry, the Janus monolayer has
distinctive behaviors by applying external electric field  $E$ with the same magnitude
but different directions ($z$ or $-z$). Our results  reveal that the
  Rashba energy ($E_R$) and  Rashba constant ($\alpha_R$) are increased
by the positive $E$, while a negative $E$  suppresses
the Rashba splitting to disappear, and then appears again. In a certain $E$ region (0.15 $\mathrm{V/\AA}$ to 0.25 $\mathrm{V/\AA}$),  switching  Rashba spin-splitting can be achieved   by only reversing  electric-field direction. Besides,  the   piezoelectric strain coefficients $d_{11}$  and $d_{31}$ (5.52 pm/V and -0.41 pm/V) are predicted, which  are higher than or  compared with those of many 2D materials. By piezoelectric effect, the strain can also be used to tune Rashba spin-splitting of RbKNaBi. In Janus RbKNaBi monolayer, the combination of piezoelectricity and Rashba spin-splitting with topological insulating phase is pregnant to promote the integration of fantastic physical phenomenons. Moreover, a possible spintronic device is proposed to  realize the function of spintronic switch. Our proposed manipulation of the  Rashba spin-splitting  may make a special contribution to semiconductor spintronics.

\end{abstract}
\keywords{Electric field, Rashba spin-splitting, Piezoelectricity  ~~~~~~~~~~~~~Email:sandongyuwang@163.com}

\maketitle

\section{Introduction}
Spintronics has attracted considerable attention, which can transmit information using spins rather than
charges by manipulating the spin degree of freedom of electrons\cite{x1}.
The semiconductor spintronic devices are compatible with the integration technologies of conventional semiconductor nanoelectronic devices.
Generating spin currents is necessary for semiconductor spintronic devices, which  can be achieved by
spin-orbit coupling (SOC)\cite{x2}. The SOC-induced spin splitting exists in noncentrosymmetric structures, mainly including
 two types: the Rashba effect induced by the structure
inversion asymmetry\cite{x3,x4} and the Dresselhaus effect
induced by the bulk inversion asymmetry\cite{x2}. Here, we concentrate on Rashba spin-splitting in two-dimensional (2D) materials.

When an electron moves across an electric field $\vec{E}$, it
experiences an effective magnetic field $\vec{B}_{eff}$ $\sim$ $\vec{E} \times \vec{p}$/$mc^2$ in its rest-frame. The effective magnetic field can induce a momentum-dependent Zeeman energy $H_{SOC}$ $\sim$ $\mu_B$($\vec{E} \times \vec{p}$)$\cdot$$\vec{\sigma}$/$mc^2$.
In crystals, the electric field $\vec{E}$ is  the gradient of the crystal potential. For 2D Janus materials, there is a built-in electric field $\vec{E}$ along the $\vec{z}$ ($\vec{E}=E_z\vec{z}$), and the spin
degeneracy of the energy spectrum is lifted, which can be described by the Rashba Hamiltonian\cite{x3,x4}:
 \begin{equation}\label{d-d}
H_R=\frac{\alpha_R}{\hbar}(\vec{z}\times\vec{p})\cdot\vec{\sigma}
 \end{equation}
where $\alpha_R$ is proportional to $E$.  The spin
splitting  of the energy spectrum  is  called Rashba spin-splitting. The Rashba semiconductors have been predicted in many 2D Janus materials\cite{x5,x6,x7,x8,x9,x10}

\begin{figure*}
  \includegraphics[width=14cm]{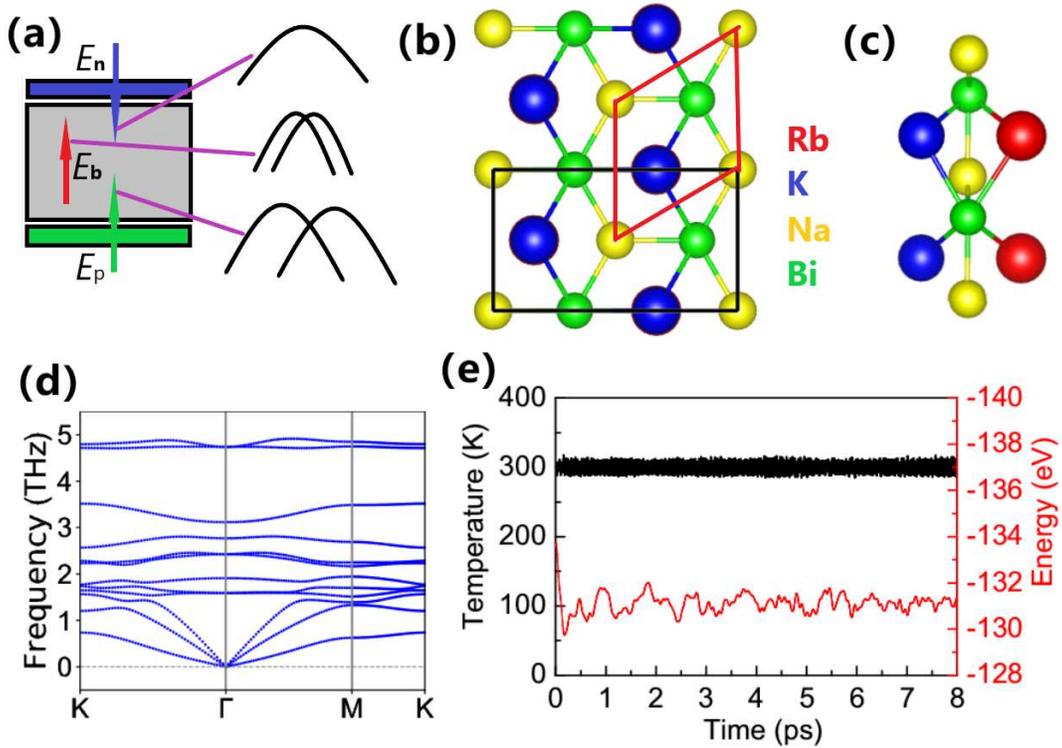}
  \caption{(Color online)(a):a 2D Janus material  with a built-in electric field $E_b$ possesses  Rashba spin-splitting. When an appropriate positive electric field $E_p$ along $z$ direction is applied, the Rashba spin-splitting  is enhanced. However, the applied electric field is reversed ($\vec{E_n}$=-$\vec{E_p}$), and  the  Rashba spin-splitting will disappear. For Janus monolayer RbKNaBi,  (b) top view and (c) side view of crystal structure, and   the   primitive cell (rectangle supercell)  is marked  by  red (black)  frames;  (d):The phonon band dispersions; (e):The temperature and total energy fluctuations as a function of simulation time at 300 K. }\label{st}
\end{figure*}

It is a natural idea to turn  on and off Rashba spin-splitting through an external electric field. As shown \autoref{st} (a), a 2D Janus material  with a built-in electric field $E_b$ possesses  Rashba spin-splitting. By applying  an appropriate positive electric field $E_p$ along $z$ direction, the Rashba spin-splitting  is enhanced. However, when the applied electric field is reversed ($\vec{E_n}$=-$\vec{E_p}$), the  Rashba spin-splitting will disappear.
In experiment, the external electric field can be as high as 0.3 $\mathrm{V/\AA}$\cite{ee}.  To accomplish our proposal practically,  a  2D Janus material should meet these conditions: (1) there is a Rashba spin-splitting, and it's better to be at the $\Gamma$ point; (2) the electronegativity difference between the bottom and top atoms is small, which will induce small built-in electric field; (3)
there are heavy elements, which may produce observable Rashba spin-splitting and remedy for small built-in electric field.

\begin{figure}
  \includegraphics[width=7cm]{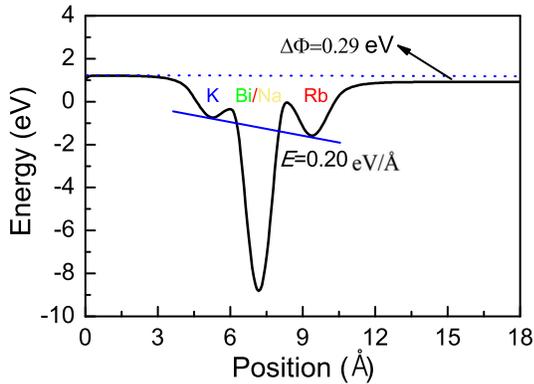}
  \caption{(Color online)For RbKNaBi monolayer, the planar averaged electrostatic potential energy variation along $z$. $\Delta\Phi$ is the potential energy difference across the layer. $E$ means built-in electric field. }\label{vot}
\end{figure}

The MoSSe is a representative 2D Janus material, which has been  synthesized experimentally\cite{e2,e1}. However, the magnitude
of the inherent electric field of MoSSe is 0.856 $\mathrm{V/\AA}$\cite{y11}, which is beyond the available experimental size of  electric field (0.3 $\mathrm{V/\AA}$)\cite{ee}.
It has been proved that the Rashba spin-splitting of MoSSe  still maintains by applying  the external electric field of -0.5 $\mathrm{V/\AA}$\cite{y11-1}.
Recently, the large-gap QSHIs $\mathrm{K_2NaBi}$ and $\mathrm{Rb_2NaBi}$
monolayers are predicted with $\mathrm{Na_3Bi}$-like crystal structure\cite{y11-2}, and they have sandwich structures.   The Na atoms are connected with Bi atoms forming
a graphene-like sheet in the ab plane, while the K/Rb atoms are above and below NaBi layer.
In this work, we construct a Janus RbKNaBi material by   replacing one of two K/Rb layers with Rb/K  atoms in monolayer  $\mathrm{K_2NaBi}$/$\mathrm{Rb_2NaBi}$.
The artificial RbKNaBi meets the three conditions mentioned above to switch  Rashba spin-splitting  by reversing  electric-field direction.
By the first-principle calculations,  our idea   is illustrated in Janus monolayer RbKNaBi as a QSHI.
It is found that  the increasing positive external electric field can enhance
  Rashba energy ($E_R$) and  Rashba constant ($\alpha_R$). However,  the increasing negative external electric field firstly quenches Rashba spin-splitting, and then recover it again. In a certain electric field region, simply reversing  electric-field direction can switch  Rashba spin-splitting.
Finally,  the   piezoelectric properties of RbKNaBi are studied, and the strain-induced electric field  by piezoelectric effect can be used to tune Rashba spin-splitting of RbKNaBi.  Our proposed manipulation of the  Rashba spin-splitting can be used for future spintronic
devices.

The rest of the paper is organized as follows. In the next
section, we shall give our computational details and methods.
 In  the next few sections,  we shall present crystal Structure and structural Stability, electronic structures along with electric field effects and piezoelectric properties of Janus monolayer RbKNaBi. Finally, we shall give our discussion and conclusion.

\begin{figure*}
  \includegraphics[width=12cm]{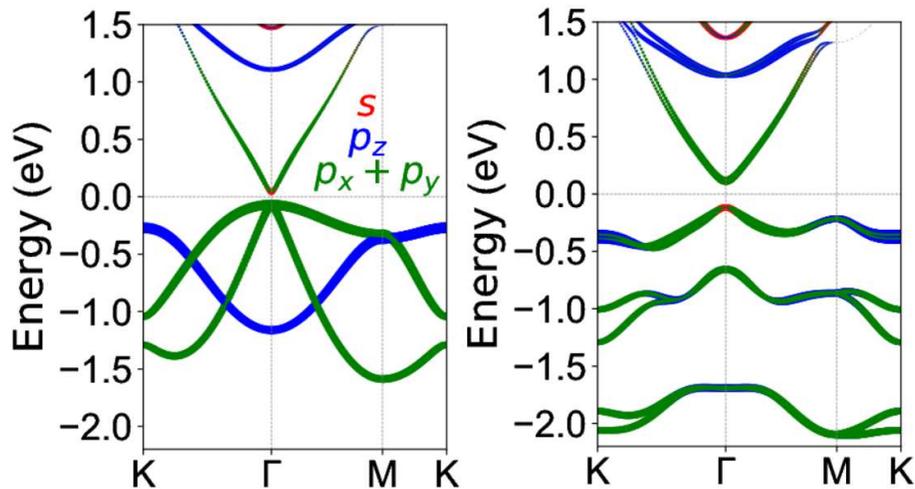}
  \caption{(Color online)  The Bi-$s$, $p_x+p_y$ and $p_z$ projected band structures of Janus monolayer RbKNaBi from GGA and GGA+SOC, respectively. The band inversion
induced by SOC can be clearly observed at the $\Gamma$ point.}\label{band}
\end{figure*}
\begin{figure}
  \includegraphics[width=7cm]{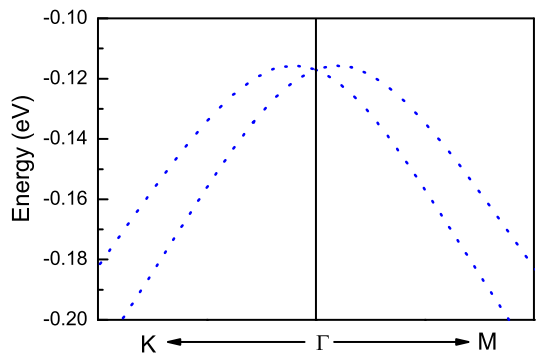}
  \caption{(Color online) The enlarged valence bands centered at the $\Gamma$ point  near the Fermi level for RbKNaBi. }\label{ra-band}
\end{figure}

\section{Computational detail}
Within density functional theory (DFT)\cite{1},   the first-principles calculations are carried out using  the projected augmented wave
(PAW) method with a kinetic cutoff energy of 500 eV, as implemented in
the  Vienna ab initio simulation package (VASP)\cite{pv1,pv2,pv3}.  The  generalized gradient approximation (GGA) of Perdew, Burke and  Ernzerhof\cite{pbe} is adopted as the exchange-correlation potential.
 The total energy  convergence criterion of $10^{-8}$ eV and residual force of  less than 0.0001 $\mathrm{eV.{\AA}^{-1}}$ on each atom are set to obtain accurate results. A vacuum spacing of larger than 16 $\mathrm{{\AA}}$ along the $z$ direction is included to avoid interactions
between two neighboring images. The SOC
is incorporated for  band structure
calculations.

The phonon dispersions are performed
using a finite difference approach  with a supercell
of 5$\times$5$\times$1, as implemented in Phonopy code\cite{pv5}.  We calculate the elastic stiffness tensor $C_{ij}$ and  piezoelectric stress coefficients $e_{ij}$ by using  strain-stress relationship (SSR) and density functional perturbation theory (DFPT) method\cite{pv6}, respectively.
The 2D elastic coefficients $C^{2D}_{ij}$
 and   piezoelectric stress coefficients $e^{2D}_{ij}$
have been renormalized by the the length of unit cell along $z$ direction ($L_z$):  $C^{2D}_{ij}$=$L_z$$C^{3D}_{ij}$ and $e^{2D}_{ij}$=$L_z$$e^{3D}_{ij}$.
A $\Gamma$-centered 12 $\times$12$\times$1 k-point meshes   in  the Brillouin zone (BZ)  is adopted to calculate  $C_{ij}$ and electronic structures, and a 8$\times$12$\times$1 Monkhorst-Pack k-point meshes for
$e_{ij}$. The WannierTools code\cite{w1} is used to investigate topological properties of RbKNaBi, based on the tight-binding Hamiltonians constructed from maximally localized Wannier functions, as  implemented in Wannier90 code\cite{w2}.
The PYPROCAR code is used to obtain the constant energy contour plots of the spin
texture\cite{py}.

\section{Crystal Structure and Structural Stability}
As shown in \autoref{st} (b) and (c), the RbKNaBi monolayer shares the honeycomb crystal with the space group of $P3m1$ (No.156).
The Na, K and Rb atoms are all bonded to the surrounding three Bi atoms, and Na and Bi atoms form a layer, while K and Rb atoms form two more layers.
The symmetry of RbKNaBi is lower than that of  $\mathrm{K_2NaBi}$/$\mathrm{Rb_2NaBi}$ with the space group of $P\bar{6}m2$ (No.187). The RbKNaBi and $\mathrm{K_2NaBi}$/$\mathrm{Rb_2NaBi}$  all lack the spatial inversion symmetry. Besides,  the RbKNaBi also lacks  horizontal mirror symmetry, which will induce out-of-plane piezoelectricity and Rashba effect.
The  Janus monolayer  RbKNaBi  can be constructed  by  replacing one of two K/Rb layers with Rb/K   atoms in monolayer $\mathrm{K_2NaBi}$/$\mathrm{Rb_2NaBi}$. The optimized  lattice parameters $a$ of RbKNaBi  is 5.587  $\mathrm{{\AA}}$, which is between ones of $\mathrm{K_2NaBi}$ (5.548  $\mathrm{{\AA}}$) and $\mathrm{Rb_2NaBi}$ (5.629  $\mathrm{{\AA}}$).

\begin{figure*}
  \includegraphics[width=12cm]{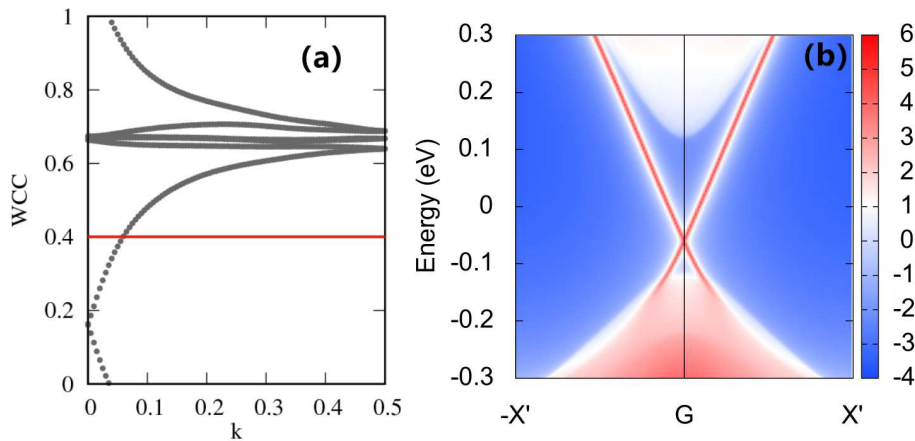}
  \caption{(Color online) For RbKNaBi monolayer, (a): calculated WCC evolutions based on the GGA+SOC electronic structure,  indicating a nontrivial $Z_2$ invariant ($Z_2$=1). (b): the projected edge spectra  with a bulk electronic structure, showing a nontrivial metallic edge state. }\label{top}
\end{figure*}

The phonon spectra,  ab initio molecular dynamics (AIMD)
simulations  and elastic constants $C_{ij}$  are calculated to confirm the stability of RbKNaBi.
As shown in \autoref{st} (d),  no imaginary vibrational frequency  can be observed, which clearly suggests that the RbKNaBi is
dynamically stable. Moreover,  both
linear and flexural modes  can be observed  around the $\Gamma$ point, which shares the general features of 2D materials\cite{r1,r2}.
The AIMD simulation is performed  by using the canonical (NVT) ensemble with a 4$\times$4$\times$1 supercell at 300 K for 8 ps
with a time step of 1 fs.  As shown in \autoref{st} (e), the total energy and temperature
fluctuate in the simulation time,  but the average energy and temperature remain
almost invariant. The final atomic configuration shows no obvious distortion of the
geometric structure after 8000 steps of AIMD simulation,  which
 can allow returning to its
initiating structure by optimizing this final configuration. These results indicate its thermal  stability of RbKNaBi at room temperature.

Due to $P3m1$ symmetry,  the 2D  elastic tensor with using Voigt notation can be reduced into:
\begin{equation}\label{pe1-4}
   C=\left(
    \begin{array}{ccc}
      C_{11} & C_{12} & 0 \\
     C_{12} & C_{11} &0 \\
      0 & 0 & (C_{11}-C_{12})/2 \\
    \end{array}
  \right)
\end{equation}
The calculated two  independent elastic
constants  $C_{11}$=18.40 $\mathrm{Nm^{-1}}$ and $C_{12}$=5.01 $\mathrm{Nm^{-1}}$, which meet the  Born  criteria of  mechanical stability \cite{ela}:
 $C_{11}>0$ and  $C_{11}-C_{12}>0$, confirming  its mechanical stability.
The shear modulus $G^{2D}$ equals to  $C_{66}$$=$($C_{11}$-$C_{12}$)/2 (6.70 $\mathrm{Nm^{-1}}$).
We calculate the direction-dependent Young's modulus $C_{2D}(\theta)$\cite{ela1}:
\begin{equation}\label{c2d}
C_{2D}(\theta)=\frac{C_{11}C_{22}-C_{12}^2}{C_{11}sin^4\theta+Asin^2\theta cos^2\theta+C_{22}cos^4\theta}
\end{equation}
where $A=(C_{11}C_{22}-C_{12}^2)/C_{66}-2C_{12}$.  Calculated results show that the RbKNaBi is  mechanically isotropic due to $P3m1$ symmetry. The  $C_{2D}$  of RbKNaBi is 17.04 $\mathrm{Nm^{-1}}$, which is between ones of $\mathrm{K_2NaBi}$ ( 18.83 $\mathrm{Nm^{-1}}$) and $\mathrm{Rb_2NaBi}$ (16.44 $\mathrm{Nm^{-1}}$)\cite{y11-2}.
The RbKNaBi monolayer demonstrates
larger mechanical flexibility than those of other well-known
2D materials (graphene  and $\mathrm{MoS_2}$)\cite{q5-1,q5-1-1}. It is found that the Poisson's ratio $\nu_{2D}(\theta)$ is independent of direction, and can be simply expressed as:
 \begin{equation}\label{e1}
\nu_{2D}=\frac{C_{12}}{C_{11}}
\end{equation}
 The calculated  $\nu_{2D}$  is 0.272.

Intrinsic polar electric field  is responsible for the
emergence of out-of-plane piezoelectricity and Rashba effect.
The difference in atomic size and electronegativity of K and Rb atoms leads to inequivalent Bi-K and Bi-Rb bond lengths (3.727 and 3.855 $\mathrm{{\AA}}$), and K-Bi-K  and Rb-Bi-Rb  bond angles (97.12 and 92.90 $^{\circ}$), giving rise to a net electric field pointing from the K layer
to the Rb layer.  To identify the inherent electric field further, the planar  average of the
electrostatic potential  energy is plotted in \autoref{vot}. The mirror asymmetry produces  an electrostatic potential gradient ($\Delta\Phi$) of about 0.29 eV, which is related to the work function change of the structure.
 The
magnitude of the net vertical electric field is estimated to be 0.20 $\mathrm{eV/{\AA}}$, which is  determined by the
slope of the plane-averaged electrostatic potential between K and Rb atoms' minima. The predicted net vertical electric field
 is smaller than that of Janus MoSSe (0.856 $\mathrm{eV/{\AA}}$)\cite{y11}, implying a
weak vertical polarization. The small built-in electric field makes for switching  Rashba spin-splitting  by reversing  electric-field direction.
\begin{figure*}
  \includegraphics[width=15cm]{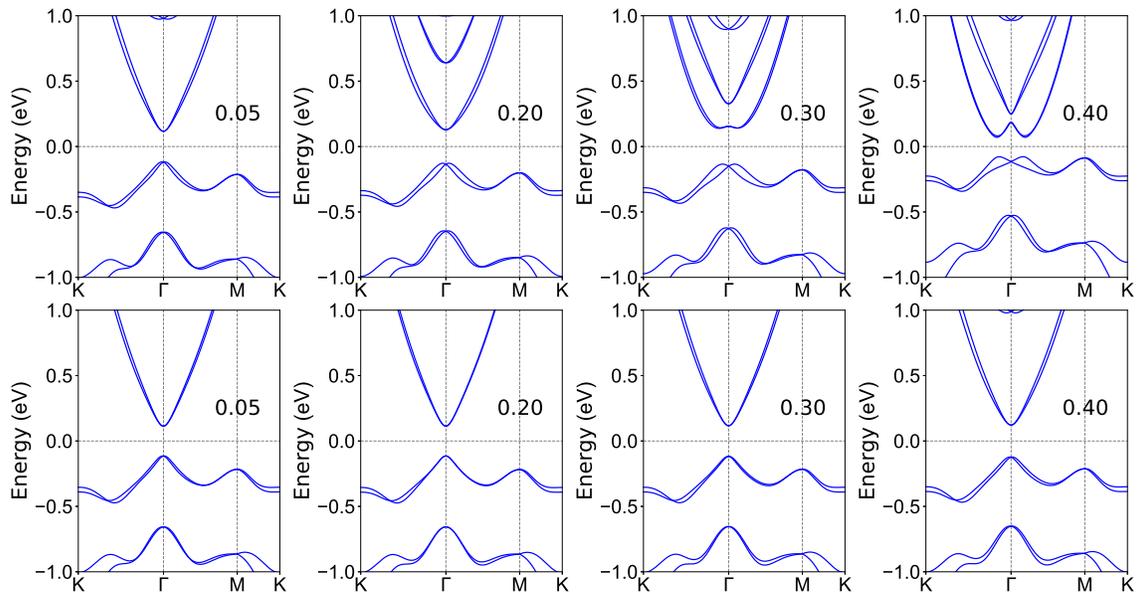}
  \caption{(Color online)For RbKNaBi monolayer, the GGA+SOC energy band structures at representative $E$=0.05, 0.20, 0.30 and 0.40 $\mathrm{V/\AA}$ with positive (Top plate) and negative (Bottom plate) electric field. }\label{band-1}
\end{figure*}
\begin{figure}
  \includegraphics[width=8cm]{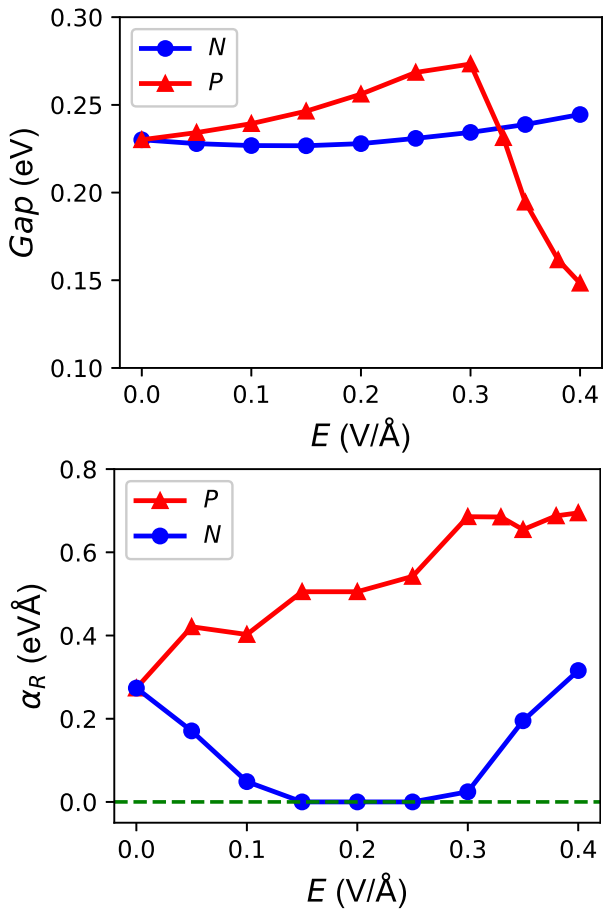}
  \caption{(Color online) For RbKNaBi monolayer, the GGA+SOC gap and  Rashba constant ($\alpha_R$) as a function of electric field $E$ with positive ($P$) and negative ($N$) electric field. }\label{ar}
\end{figure}

\begin{figure*}
  \includegraphics[width=15cm]{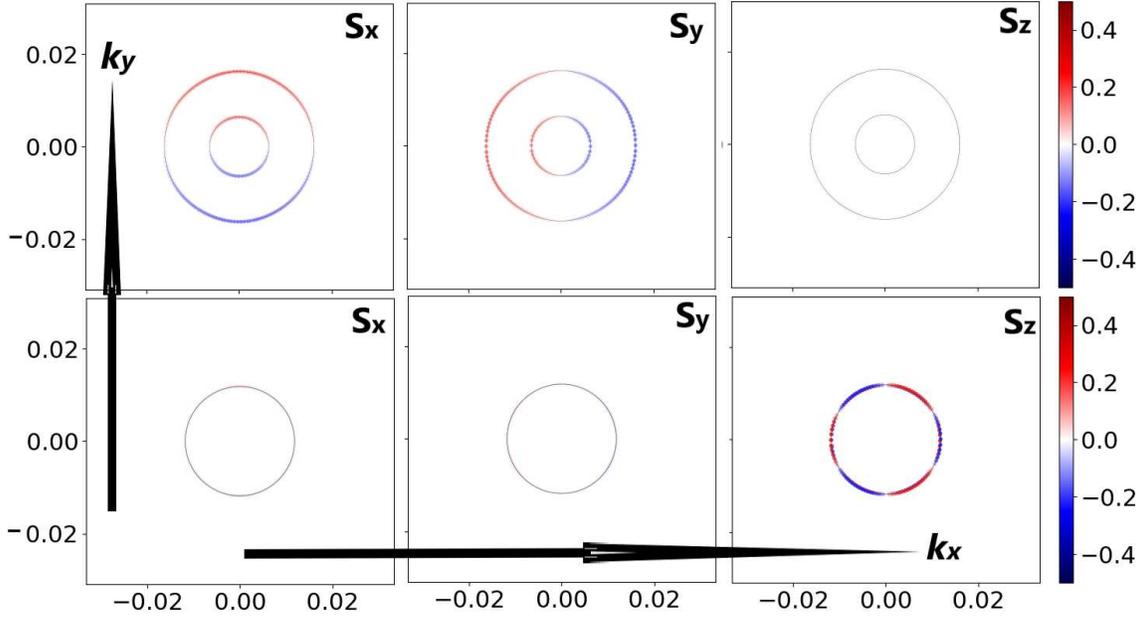}
  \caption{(Color online)For RbKNaBi monolayer, the spin projected constant energy contour plots of spin texture calculated in a $k_x$-$k_y$ plane centered at the $\Gamma$ point at an energy surface 0.16 eV below  the Fermi level with  positive (Top plate) and negative (Bottom plate) electric field $E$=0.20 $\mathrm{V/\AA}$. In the color scale, red means spin-up states while blue means spin-down states.  }\label{rash}
\end{figure*}

\begin{figure}
  \includegraphics[width=8cm]{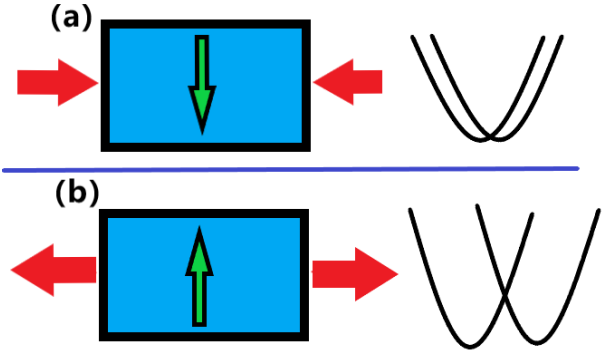}
  \caption{(Color online) Schematic of biaxial in-plane strain tuned Rashba spin-splitting: the red arrows represent strain, while the green arrows represent electric field induced by piezoelectric effect; (a) the compressive strain induce negative electric field, which can reduce Rashba spin-splitting. (b) the tensile strain induce positive electric field, which can enhance Rashba spin-splitting. }\label{yd}
\end{figure}

\begin{table*}
\centering \caption{For monolayer $\mathrm{K_2NaBi}$, RbKNaBi and $\mathrm{Rb_2NaBi}$, the elastic constants $C_{ij}$ ($\mathrm{Nm^{-1}}$), piezoelectric stress coefficient $e_{ij}$ along with electronic part $e_{ij_e}$ and  ionic part $e_{ij_i}$ ($10^{-10}$ C/m), and piezoelectric strain coefficient $d_{ij}$ (pm/V). }\label{tab0}
\begin{tabular*}{0.96\textwidth}{@{\extracolsep{\fill}}ccccccccccc}
  \hline\hline
Name& $C_{11}$&	$C_{12}$&		$e_{11_e}$&	$e_{11_i}$&	$e_{11}$&	$e_{31_e}$&	$e_{31_i}$&	$e_{31}$&	$d_{11}$&	$d_{31}$\\\hline
$\mathrm{K_2NaBi}$&19.46&	5.29&	0.15 &	0.66&	0.81&	--&	--&	--&	5.69 &	--\\\hline
RbKNaBi&18.41&	5.02&	0.23&	0.51&	0.74&	-0.071 &	-0.024&	-0.095&	5.52&	-0.41\\\hline
$\mathrm{Rb_2NaBi}$&  17.08&	4.51&	0.293&	0.348&	0.641&	--&	--&	--&5.10&	--	\\\hline\hline
\end{tabular*}
\end{table*}

\section{electronic structures and electric field effects}
Monolayer $\mathrm{K_2NaBi}$  and $\mathrm{Rb_2NaBi}$  are predicted are large-gap QSHIs\cite{y11-2}.  The  nontrivial topological properties may be broken by building Janus structure from a QSHI, for example  Janus  $\mathrm{SrAlGaSe_4}$ and MoSSe as derivatives of QSHIs $\mathrm{SrGa_2Se_4}$ and 1$T'$-$\mathrm{MoS_2}$\cite{gsd4,gsd4-1}.
Next, we investigate the electronic properties of RbKNaBi from GGA and GGA+SOC, and the   Bi-$s$, $p_x+p_y$ and $p_z$ projected
band structures are plotted in \autoref{band}. The GGA results show that RbKNaBi  has a direct
band gap of 0.110 eV  with valence band maximum (VBM) and conduction band minimum (CBM) at the $\Gamma$ point.
However, the GGA+SOC results show an indirect  band gap of 0.229 eV, and the VBM has a slight deviation from the $\Gamma$ point.
The calculated gap within SOC is close to those of $\mathrm{K_2NaBi}$  (0.18 eV) and $\mathrm{Rb_2NaBi}$ (0.22 eV)\cite{y11-2}.  Without considering SOC, Bi-$p_x+p_y$ orbitals contribute to the VBM, while the CBM is dominated by Bi-$s$
states. When including SOC,  an inversion of Bi-$s$ and Bi-$p_x+p_y$
states  can be observed in the band structures,  suggesting that monolayer RbKNaBi is a QSHI. On the other hand,
the SOC can  lift the degenerate spins  due to relativistic effects.  For Janus structures, the mirror asymmetry can produce built-in electric field,
leading to Rashba spin-splitting (see \autoref{ra-band}). The strength of the Rashba effect can be quantized by the Rashba energy ($E_R$) and  Rashba constant ($\alpha_R$), and the $\alpha_R=2E_R/k_o$ with $k_o$ for the Rashba momentum (see FIG.1 of electronic supplementary information (ESI)). The calculated $E_R$= 1.3 meV and $\alpha_R$=0.274 $\mathrm{eV{\AA}}$.

To further confirm nontrivial topological properties of RbKNaBi, we calculate the $Z_2$ topological
invariant. For a material with  inversion symmetry, we can calculate $Z_2$ topological
invariant by the product of parities of all
occupied states at four time-reversal-invariant-momentum
 points in the 2D BZ.
 However,  for RbKNaBi with broken spatial inversion symmetry,   the $Z_2$  can be obtained  by the calculation of
Wannier charge center (WCC)\cite{wcc}.
If $Z_2$ equals 1, a material is  a topologically nontrivial, while $Z_2$$=$0 means  trivial state.
 As plotted in \autoref{top} (a), the evolution lines of WCC  cross an odd number of times by an arbitrary reference line, giving rise to $Z_2$=1, which indicates
that monolayer RbKNaBi is a QSHI.  The projected edge spectra  along the [100]
direction is calculated, and the local
density of states (LDOS) is shown  in \autoref{top} (b). It is clearly seen that there is a single pair of helical edge states in the bulk bandgap.
Remarkably, a
sizeable bulk gap (229 meV)  makes for observing the room-temperature quantum spin Hall (QSH) effect, because the large gap can stabilize the helical edge states against the interference of the thermally activated carriers.

A perpendicular electric field is used to explore the manipulation of the Rashba spin
splitting of RbKNaBi. There are different atomic species on its upper and lower facets for Janus RbKNaBi,  implying that applying $+z$ and $-z$ directional electric field are not equivalent. The energy band structures as a function of electric field  $E$ are calculated by using GGA+SOC. Some representative energy band structures are plotted in \autoref{band-1}, and the energy band gap vs $E$ are shown in \autoref{ar}.
In considered positive $E$ range, the gap firstly increases with increasing $E$, and then has a sudden drop at $E$=0.30 $\mathrm{V/\AA}$. However, the negative $E$ has small effects on gap. In considered both positive and negative $E$ ranges, the RbKNaBi is always a QSHI, and the calculated WCC evolutions at representative  positive  and negative $E$=0.20 $\mathrm{V/\AA}$ are plotted in FIG.2 of ESI.

Next, we mainly investigate electric field effects on Rashba spin
splitting of RbKNaBi. The  calculated $E_R$ as a function of both positive and negative $E$ are presented in FIG.3 of ESI.
For positive case, the $E_R$ increases with increasing $E$. However, for negative situation,  the  $E_R$ firstly decreases with increasing $E$, and become almost zero in a certain $E$ region, and then increases. This is due to a competition between  the external electric field and internal electric field.
To clearly see the disappeared Rashba spin
splitting, the  enlarged valence bands centered at the $\Gamma$ point  near the Fermi level at representative  positive  and negative $E$=0.20 $\mathrm{V/\AA}$ are plotted in FIG.4 of ESI.  The  calculated $\alpha_R$ vs  $E$ along both positive and negative directions are presented in \autoref{ar}.
The overall trend of $\alpha_R$ vs $R$ is consistent with that of $E_R$ vs $E$. It is clearly seen that the Rashba spin
splitting of RbKNaBi is nonexistent, when $E$ is between negative 0.15 $\mathrm{V/\AA}$ and 0.25 $\mathrm{V/\AA}$.  In this region, the Rashba spin
splitting  will be opened by  reversing  electric field direction (from negative $E$ to positive $E$). This switch operation  can expand the range of possible applications in future spintronic technology.

Besides , the constant energy (at an energy surface 0.16 eV below
the Fermi level) spin-resolved 2D contours for the spin
texture centered at the $\Gamma$ point are plotted in \autoref{rash} under both positive and negative $E$$=$0.2 eV/$\mathrm{{\AA}}$.  For positive case,
the  $S_y$ components have  a 90$^{\circ}$  of rotation as compared to $S_x$ counterpart.
The concentric spin-texture circles mean the purely 2D Rashba spin-splitting (an isotropic splitting).
 Only in-plane $S_x$ and $S_y$ spin components are
present in the Rashba spin-splitting bands, while the out-of-plane $S_z$ component is non-existent, which  further confirms that the
spin splitting under  positive  $E$$=$0.2 eV/$\mathrm{{\AA}}$ has an isotropic
2D Rashba nature. For negative situation, no Rashba spin-splitting bands can be observed.  Only out-of-plane $S_z$ spin component is
present, while in-plane $S_x$ and $S_y$ components vanish. It is clearly seen that the out-of-plane $S_z$ spin component has three-fold rotational symmetry.

\section{Piezoelectric properties}
 When a  strain or stress is applied on a non-centrosymmetric material, the  electric
dipole moments can be induced and produce electricity, called  piezoelectric effect.
The piezoelectric response of a material can be described by third-rank piezoelectric stress tensor  $e_{ijk}$ and strain tensor $d_{ijk}$.  The relaxed piezoelectric tensors ($e_{ijk}$ and $d_{ijk}$) include the ionic and electronic contributions:
 \begin{equation}\label{pe0}
      e_{ijk}=\frac{\partial P_i}{\partial \varepsilon_{jk}}=e_{ijk}^{elc}+e_{ijk}^{ion}
 \end{equation}
and
 \begin{equation}\label{pe0-1}
   d_{ijk}=\frac{\partial P_i}{\partial \sigma_{jk}}=d_{ijk}^{elc}+d_{ijk}^{ion}
 \end{equation}
in which $P_i$, $\varepsilon_{jk}$ and $\sigma_{jk}$ are polarization vector, strain and stress, respectively, and the superscripts $elc$ and $ion$ is used to denote electronic and ionic contributions. The  $e_{ijk}^{elc}$ and $d_{ijk}^{elc}$  are also called clamped-ion  piezoelectric coefficients, while $e_{ijk}$ and $d_{ijk}$ mean  relax-ion cases.
The $e_{ijk}$ is related with  $d_{ijk}$  by elastic tensor $C_{mnjk}$:
 \begin{equation}\label{pe0-1-1}
    e_{ijk}=\frac{\partial P_i}{\partial \varepsilon_{jk}}=\frac{\partial P_i}{\partial \sigma_{mn}}.\frac{\partial \sigma_{mn}}{\partial\varepsilon_{jk}}=d_{imn}C_{mnjk}
 \end{equation}

With respect to $\mathrm{K_2NaBi}$  and $\mathrm{Rb_2NaBi}$, for RbKNaBi, the introduction of Janus structure results in a lower degree of $C_{3v}$ symmetry, and both the in-plane and
out-of-plane piezoelectric effects are allowed, when a uniaxial in-plane  strain is applied.  By using  Voigt notation,  the 2D  piezoelectric stress and  strain  tensors  can be expressed as\cite{q5,q5-11}:
 \begin{equation}\label{pe1-1}
 e=\left(
    \begin{array}{ccc}
      e_{11} & -e_{11} & 0 \\
     0 & 0 & -e_{11} \\
      e_{31} & e_{31} & 0 \\
    \end{array}
  \right)
    \end{equation}

  \begin{equation}\label{pe1-2}
  d= \left(
    \begin{array}{ccc}
      d_{11} & -d_{11} & 0 \\
      0 & 0 & -2d_{11} \\
      d_{31} & d_{31} &0 \\
    \end{array}
  \right)
\end{equation}
    It is found that  only out-of-plane piezoelectric response can exist ($e_{11}$/$d_{11}$=0, but $e_{31}$/$d_{31}$$\neq$0), when applying  a biaxial in-plane strain. In other words,  the pure out-of-plane piezoelectric response can be realized by imposed biaxial strain.  Here, the two independent $d_{11}$ and $d_{31}$ can be derived by $e_{ik}=d_{ij}C_{jk}$:
\begin{equation}\label{pe2}
    d_{11}=\frac{e_{11}}{C_{11}-C_{12}}~~~and~~~d_{31}=\frac{e_{31}}{C_{11}+C_{12}}
\end{equation}

We use the orthorhombic supercell (see  \autoref{st} (b))  to calculate the  $e_{11}$/$e_{31}$ of RbKNaBi.
The predicted  $e_{11}$/$e_{31}$ is 0.74$\times$$10^{-10}$/-0.095$\times$$10^{-10}$ C/m  with ionic part 0.51$\times$$10^{-10}$/-0.024$\times$$10^{-10}$ C/m  and electronic part 0.23$\times$$10^{-10}$/-0.071$\times$$10^{-10}$ C/m.
For both  $e_{11}$ and $e_{31}$, the electronic and ionic parts have superposed contributions.  Based on \autoref{pe2}, the predicted  $d_{11}$/$d_{31}$  is  5.52/-0.41 pm/V. The predicted $d_{11}$ and $d_{31}$ (absolute value) are higher than or  compared with those of familiar 2D materials\cite{q5,q5-11}.
The $E$ along $z$ direction can also be induced with a biaxial in-plane strain  by piezoelectric effect, and then tune  Rashba spin-splitting of RbKNaBi.
When an in-plane vibration is applied to RbKNaBi, the Rashba effect will be modulated periodically. As shown in \autoref{yd} (a), the compressive strain induce negative electric field, which can reduce Rashba spin-splitting. However, \autoref{yd} (b) shows that the tensile strain induce positive electric field, which can enhance Rashba spin-splitting.

The monolayer $\mathrm{K_2NaBi}$  and $\mathrm{Rb_2NaBi}$ possess horizontal mirror symmetry, and only in-plane $d_{11}$ can exist, when a uniaxial in-plane strain is applied.For comparison, we also calculate the piezoelectric coefficients of  $\mathrm{K_2NaBi}$  and $\mathrm{Rb_2NaBi}$, and the related data are listed in \autoref{tab0}. It is found that the $d_{11}$ of the three monolayers are very close, and the  $d_{11}$ of RbKNaBi
monolayer fall in between those of the $\mathrm{K_2NaBi}$  and $\mathrm{Rb_2NaBi}$ monolayers, as expected.
The coexistence of piezoelectricity  and  nontrivial topological insulating phase, namely piezoelectric quantum spin Hall insulator (PQSHI), has potential advantages toward the development of high-speed and dissipationless electronic devices. Combining their nontrivial electronic structures, $\mathrm{K_2NaBi}$, $\mathrm{Rb_2NaBi}$ and  RbKNaBi all are PQSHIs.

\begin{figure}
  \includegraphics[width=7cm]{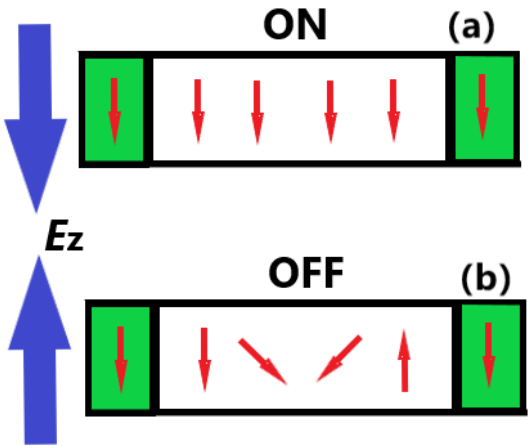}
  \caption{(Color online) Schematic of spintronic devices: the red arrows represent spin, while the blue arrows represent electric field.  }\label{qj}
\end{figure}

\section{Discussion and Conclusion}
 Some suggestions on experimental synthesis aspects of RbKNaBi  are discussed.  Firstly, the dynamically stable bulk
compounds $\mathrm{K_2NaBi}$  and $\mathrm{Rb_2NaBi}$ are  thermodynamically stable against
disproportionation into the competing phases\cite{bu}. Secondly, by exfoliating these $\mathrm{Na_3Bi}$-like alkali bismide three-dimensional Dirac
semimetals $\mathrm{K_2NaBi}$  and $\mathrm{Rb_2NaBi}$, their monolayer structures are also dynamically,  thermally  and  mechanically stable\cite{y11-2}.
Finally, similar to Janus monolayer MoSSe from $\mathrm{MoS_2}$\cite{e1,e2}, the  RbKNaBi   can be synthesized experimentally with similar experimental techniques based on $\mathrm{K_2NaBi}$  or $\mathrm{Rb_2NaBi}$ monolayer.

The SOC effects (Rashba spin-splitting) can be reversibly turned on and off  by reversing  electric-field direction, which  provides a high-speed switch for the subsequent development of spin devices to control the passage of electrons (see \autoref{qj}). For example, two ferromagnetic electrodes in the same direction are set for the component. With an appropriate negative applied electric field, the injected electrons pass through the RbKNaBi channel at a high speed and keep the spin orientation unchanged (\autoref{qj} (a)). When the external electric field direction is reversed, the spin of  electrons in the channel will  rotate under the effect of SOC, and will be blocked by the derived electrode (\autoref{qj} (b)). This possible  spintronic device realizes the function of spintronic switch.

In summary, we have demonstrated that Rashba spin-splitting can be reversibly turned on and off  by reversing  electric-field direction
in RbKNaBi.   Calculated results show that the $E_R$ and  $\alpha_R$ are increased
by the positive $E$, while a negative $E$  suppresses
the Rashba splitting to disappear, and then appears again. In a certain electric field region, simply  reversing  electric-field direction can achieve switching  Rashba spin-splitting. Besides, it is proved that RbKNaBi is a QSHI and possesses excellent piezoelectric properties, which makes RbKNaBi become a multifunctional 2D material.
Our findings can inspire more works about  electric filed-tuned  switch of  Rashba spin-splitting.
%~~~~\\
%~~~~\\
%\textbf{SUPPLEMENTARY MATERIAL}
%\\
%See the supplementary material for  crystal structures; energy difference between AFM and FM and local magnetic moment of V atom as a function of $E$; the related energy band structures and Berry curvatures.
%
%
%
%
%
%~~~~\\
%~~~~\\
%\textbf{Conflicts of interest}
%\\
%There are no conflicts to declare.

\begin{acknowledgments}
This work is supported by Natural Science Basis Research Plan in Shaanxi Province of China  (2021JM-456). We are grateful to Shanxi Supercomputing Center of China, and the calculations were performed on TianHe-2.
\end{acknowledgments}

\end{document}